\documentstyle[11pt]{article}
\expandafter\ifx\csname amssym.def\endcsname\relax \else\endinput\fi
\expandafter\edef\csname amssym.def\endcsname{%
       \catcode`\noexpand\@=\the\catcode`\@\space}
\catcode`\@=11
\def\undefine#1{\let#1\undefined}
\def\newsymbol#1#2#3#4#5{\let\next@\relax
 \ifnum#2=\@ne\let\next@\msafam@\else
 \ifnum#2=\tw@\let\next@\msbfam@\fi\fi
 \mathchardef#1="#3\next@#4#5}
\def\mathhexbox@#1#2#3{\relax
 \ifmmode\mathpalette{}{\m@th\mathchar"#1#2#3}%
 \else\leavevmode\hbox{$\m@th\mathchar"#1#2#3$}\fi}
\def\hexnumber@#1{\ifcase#1 0\or 1\or 2\or 3\or 4\or 5\or 6\or 7\or 8\or
 9\or A\or B\or C\or D\or E\or F\fi}

\font\tenmsa=msam10
\font\sevenmsa=msam7
\font\fivemsa=msam5
\newfam\msafam
\textfont\msafam=\tenmsa
\scriptfont\msafam=\sevenmsa
\scriptscriptfont\msafam=\fivemsa
\edef\msafam@{\hexnumber@\msafam}
\mathchardef\dabar@"0\msafam@39
\def\dashrightarrow{\mathrel{\dabar@\dabar@\mathchar"0\msafam@4B}}
\def\dashleftarrow{\mathrel{\mathchar"0\msafam@4C\dabar@\dabar@}}

\def\ulcorner{\delimiter"4\msafam@70\msafam@70 }
\def\urcorner{\delimiter"5\msafam@71\msafam@71 }
\def\llcorner{\delimiter"4\msafam@78\msafam@78 }
\def\lrcorner{\delimiter"5\msafam@79\msafam@79 }
\def\yen{{\mathhexbox@\msafam@55}}
\def\checkmark{{\mathhexbox@\msafam@58}}
\def\circledR{{\mathhexbox@\msafam@72}}
\def\maltese{{\mathhexbox@\msafam@7A}}
\def\circledS{{\mathhexbox@\msafam@73}}

\font\tenmsb=msbm10
\font\sevenmsb=msbm7
\font\fivemsb=msbm5
\newfam\msbfam
\textfont\msbfam=\tenmsb
\scriptfont\msbfam=\sevenmsb
\scriptscriptfont\msbfam=\fivemsb
\edef\msbfam@{\hexnumber@\msbfam}
\def\Bbb#1{{\fam\msbfam\relax#1}}
\def\widehat#1{\setbox\z@\hbox{$\m@th#1$}%
 \ifdim\wd\z@>\tw@ em\mathaccent"0\msbfam@5B{#1}%
 \else\mathaccent"0362{#1}\fi}
\def\widetilde#1{\setbox\z@\hbox{$\m@th#1$}%
 \ifdim\wd\z@>\tw@ em\mathaccent"0\msbfam@5D{#1}%
 \else\mathaccent"0365{#1}\fi}
\font\teneufm=eufm10
\font\seveneufm=eufm7
\font\fiveeufm=eufm5
\newfam\eufmfam
\textfont\eufmfam=\teneufm
\scriptfont\eufmfam=\seveneufm
\scriptscriptfont\eufmfam=\fiveeufm
\def\frak#1{{\fam\eufmfam\relax#1}}

\csname amssym.def\endcsname

\makeatletter
\def\section{\@startsection {section}{1}{\z@}{-3.5ex plus -1ex minus 
 -.2ex}{2.3ex plus .2ex}{\large\sc}}
\def\subsection{\@startsection{subsection}{2}{\z@}{-3.25ex plus -1ex minus 
 -.2ex}{1.5ex plus .2ex}{\normalsize\sc}}
\makeatother

\makeatletter
\@addtoreset{equation}{section}                        

\makeatother

\newcommand{\nc}{\newcommand}
\newcommand{\rnc}{\renewcommand}

\nc{\be}{\begin{equation}}
\nc{\ee}{\end{equation}}
\nc{\bea}{\begin{eqnarray}}
\nc{\eea}{\end{eqnarray}}

\nc{\trac}[2]{{\textstyle\frac{#1}{#2}}}

\nc{\ex}[1]{\mbox{e}^{\,\textstyle#1}}

\nc{\CC}{\Bbb{C}}
\nc{\HH}{\Bbb{H}}
\nc{\PP}{\Bbb{P}}
\nc{\RR}{\Bbb{R}}
\nc{\ZZ}{\Bbb{Z}}
\nc{\II}{\Bbb{I}}
\nc{\EE}{\Bbb{E}}
\nc{\SS}{\Bbb{S}}

\rnc{\a}{\alpha}
\rnc{\b}{\beta}
\nc{\al}{\a^{l}}
\rnc{\d}{\delta}
\nc{\ga}{\gamma}
\nc{\la}{\lambda}
\nc{\lal}{\la_{l}}
\nc{\f}{\phi}
\nc{\fb}{\bar{\phi}}
\nc{\p}{\psi}

\nc{\e}{\eta}
\nc{\eb}{\bar{\eta}}
\rnc{\c}{\chi}
\nc{\eps}{\epsilon}
\rnc{\t}{\theta}
\nc{\tb}{\bar{\theta}}
\nc{\om}{\omega}

\rnc{\P}{\Psi}
\nc{\pl}{\P_{L}}
\nc{\pdr}{\P^{\dag}_{R}}
\nc{\G}[3]{\Gamma^{#1}_{\;{#2}{#3}}}
\nc{\Gd}[3]{\Gamma_{{#1}{#2}{#3}}}
\nc{\Ga}{\Gamma}

\nc{\sig}{\sigma}
\nc{\sk}{\sigma_{k}}
\nc{\sa}{\sigma_{a}}
\nc{\Bb}{\bar{B}}

\nc{\symx}{\circledS}

\nc{\Q}{\bar{Q}}
\nc{\C}{{\cal A}/{\cal G}}                
\nc{\A}[1]{{\cal A}^{#1}/{\cal G}^{#1}}  
\nc{\RC}{{\cal R}_{\C}}                 
\nc{\RM}{{\cal R}_{\M}}                
\nc{\RX}{{\cal R}_{X}}
\nc{\RY}{{\cal R}_{Y}}

\nc{\ad}{\mathop{\mbox{ad}}\nolimits}
\nc{\tr}{\mathop{\mbox{tr}}\nolimits}
\nc{\Tr}{\mathop{\mbox{Tr}}\nolimits}
\nc{\Det}{\mathop{\mbox{Det}}\nolimits}
\rnc{\det}{\mathop{\mbox{det}}\nolimits}
\nc{\rk}{\mathop{\mbox{rk}}\nolimits}
\nc{\diag}{\mbox{diag}}
\nc{\ra}{\rightarrow}
\nc{\Ra}{\Rightarrow}
\nc{\LRa}{\Leftrightarrow}
\nc{\lra}{\leftrightarrow}
\nc{\ot}{\otimes}
\rnc{\ss}{\subset}
\nc{\nul}{\noindent\underline}
\nc{\non}{\nonumber\\}
\rnc{\S}{\Sigma}
\nc{\tp}{2\pi i}
\nc{\del}{\partial}
\nc{\dbar}{\bar{\del}}
\nc{\dx}{\dot{x}}
\nc{\dxl}{\dot{x}^{\la}}
\nc{\dxm}{\dot{x}^{\mu}}
\nc{\dxn}{\dot{x}^{\nu}}
\nc{\ddx}{\ddot{x}}
\nc{\ddxm}{\ddot{x}^{\mu}}
\nc{\dxi}{\dot{\xi}}
\nc{\ddxi}{\ddot{\xi}}
\nc{\zb}{\bar{z}}
\nc{\na}{\nabla}
\nc{\nal}{\nabla_{\la}}
\nc{\nam}{\nabla_{\mu}}
\nc{\nan}{\nabla_{\nu}}
\nc{\gint}{\int \sqrt{g} d^{4}x\;}

\nc{\gz}{g^{(0)}}
\nc{\gt}{g^{(2)}}
\nc{\gf}{g^{(4)}}
\nc{\ff}{f^{(4)}}

\rnc{\lg}{\frak{g}}

\nc{\lv}{\log V_{s}}
\nc{\vs}{V_{s}}
\rnc{\ln}{\log \N}
\nc{\ls}{\ell_{s}}
\nc{\N}{{\cal N}}
\nc{\M}{{\cal M}}
\nc{\F}{{\cal F}}
\nc{\E}{{\cal E}}
\rnc{\P}{{\cal P}}
\nc{\I}{{\cal I}}
\nc{\IIt}{$\widetilde{\mbox{II}}$}
\nc{\gst}{\widetilde{g_{s}}}
\nc{\gsh}{\widehat{g_{s}}}
\nc{\lsh}{\widehat{\ls}}
\nc{\rllh}{\widehat{R_{11}}}
\nc{\lph}{\widehat{\ell_{P}}}
\nc{\mnd}{M_{Nd}}

\nc{\mat}[4]{\left(\begin{array}{cc}#1&#2\\#3&#4\end{array}\right)}
 
\nc{\r}[1]{{\mathbf{#1}}}
\nc{\rb}[1]{{\overline{\mathbf{#1}}}}

\nc{\gi}{\gamma_{i}}
\nc{\gj}{\gamma_{j}}
\nc{\adss}{AdS_{5}\times S^{5}}
\nc{\adsx}{AdS_{5}\times X^{5}}

\nc{\subs}[1]{\begin{center}%
{{\sc #1}}{\addcontentsline{toc}{subsection}{#1}}%
\end{center}}

\nc{\chap}[1]{{\clearpage}%
\begin{center}%
{\noindent\underline{\large\sc #1}}{\addcontentsline{toc}{section}{#1}}%
\end{center}%
{\vspace*{0.3cm}}}

\newcommand{\ba}{\begin{eqnarray}}
\newcommand{\ea}{\end{eqnarray}}

\begin{document}

\begin{titlepage}

\begin{flushright}
{\tt hep-th/9904179}
\end{flushright}
\vspace*{0.5in}
\begin{center}
{\LARGE{\sc 
On Subleading Contributions to the \\[.2in] AdS/CFT Trace Anomaly}}\\
\vskip .3in
{\large\sc Matthias Blau}\footnote{e-mail: mblau@ictp.trieste.it} and 
{\sc K.S.\ Narain}\footnote{e-mail: narain@ictp.trieste.it}\\
\vspace{.2in}
{\it ICTP, Strada Costiera 11, 34014 Trieste, Italy}\\
\vskip .3in
{\large\sc Edi Gava}\footnote{e-mail: gava@he.sissa.it}\\
\vspace{.2in}
{\it INFN, ICTP and SISSA, Trieste, Italy}
\end{center}

\begin{abstract}
\noindent In the context of the AdS/CFT correspondence, we perform a direct
computation in $AdS_5$ supergravity of the trace anomaly of a $d=4$,
${\cal N}=2$ SCFT. We find agreement with the field theory result up
to next to leading order in the $1/N$ expansion. In particular, the
order $N$ gravitational contribution to the anomaly is  obtained from a
Riemann tensor squared term in the 7-brane effective action deduced from
heterotic - type I duality.  We also discuss, in the AdS/CFT context,
the order $N$ corrections to the trace anomaly in $d=4$, ${\cal N}=4$
SCFTs involving SO or Sp gauge groups.

\end{abstract}
\end{titlepage}
\setcounter{footnote}{0}

\section{Introduction}

Recently, a lot of work has been done on the conjectured \cite{jm1}
AdS/CFT correspondence between string theory or M-theory compactifications 
on $AdS_{d+1}$ and $d$-dimensional conformal field theories.  In particular,
this conjecture relates \cite{gkp,ew1} correlation functions of local
operators in the conformal field theory to amplitudes in the `bulk'
string theory or M-theory, with the boundary values of the bulk fields 
interpreted as sources coupling to the operators of the `boundary'
conformal field theory. 

An example of particular interest is the conjectured equivalence
between ${\cal N}=4,d=4$ supersymmetric $SU(N)$ Yang-Mills theory
and type IIB superstring theory on $AdS_{5}\times S^{5}$ (with $N$
units of RR five-form $F^{(5)}$ flux on $S^{5}$). This duality 
identifies the complex gauge coupling constant
\be
\tau_{YM}=\frac{\theta}{2\pi} + \frac{4\pi i}{g^{2}_{YM}}
\ee
of the SYM theory with the constant expectation value 
of the type IIB string coupling 
\be
\tau_{s} = \langle C^{(0)} + i\ex{-\phi}\rangle \equiv
                    \frac{\chi}{2\pi} + \frac{i}{g_{s}}\;\;. 
\ee
The radius of $S^{5}$ (or the curvature radius of $AdS_{5}$) is
\be
L=(4\pi g_{s}N)^{1/4} \ell_{s}\;\;,
\label{L1}
\ee 
with $\ell_{s}$ the string length,
$\ell_{s}^{2}=\a'$. In terms of the 't Hooft coupling $\la = g^{2}_{YM}N$,
the dimensionless scale $L^{2}/\a'$ of string theory on $AdS_{5}\times
S^{5}$ is related to the SYM parameters by 
\be
\la^{1/2} = \frac{L^{2}}{\a'}\;\;.
\ee

Correlation functions of string theory on $\adss$ are given by a double
expansion in $g_{s}$ and $\a'/L^{2}$, which can be written as a double
expansion in terms of $1/N= 4\pi g_{s}(\a'/L^{2})^{2}$ and $\la^{-1/2}$.
For closed oriented strings this is actually an expansion in even powers
of $N$, the string theory tree-level (supergravity) contribution being
of order $N^{2}$. 

Correlation functions of the SYM theory, on the other hand, have a
$1/N$ expansion, valid when $N$ is large, $g^{2}_{YM}$ is small, and
$\la$ is kept finite (and small). For $SU(N)$ theories with adjoint fields
only, this is once again an expansion in even powers of $N$, the leading
contributions, of order $N^{2}$, coming from planar diagrams.

According to the (strong form of the) AdS/CFT correspondence, these
two theories should give rise to the same function of $\la$ at each 
order in $N$. However, as one has an expansion in terms of $\la$ (weak
't Hooft coupling) and the other in terms of $\la^{-1/2}$ (the $\a'$ or,
better, $\a'/L^{2}$, expansion of string theory), in practice this
comparison is restricted to the rather limited set of quantities which 
are $\la$-independent, such as global anomalies. 

Leading order $N^{2}$
contributions to the chiral anomalies were checked in e.g.\ 
\cite{ew1,ca}, and trace anomalies were discussed (at the linearized
level) in \cite{lt} by comparing the bulk supergravity action with the 
effective action arising from the coupling of ${\cal N}=4$ SYM to 
${\cal N}=4$ conformal supergravity.

The complete leading order `holographic Weyl anomaly' was determined 
in general in \cite{hs}. In 
particular, it was found there that the leading supergravity
contribution to the trace anomaly involves only the squares of the
Ricci tensor and Ricci
scalar of the boundary metric and not the square of the Riemann tensor
itself. This implies that conformal field theories with a
standard (product space) supergravity dual necessarily have $a=c$
to leading order in $N$, where $a$ and $c$ are the coefficients of the
Euler and Weyl terms in the standard expression 
\be
\langle T^{\mu}_{\;\mu}\rangle = -a E_{4}-cI_{4}
\ee
for the conformal anomaly. In \cite{NO2} the calculations of
\cite{hs} have been generalized to dilatonic gravity.

The check of subleading corrections in the $1/N^{2}$ expansion 
is hampered by the fact that, for closed string theory, these correspond 
to string loop corrections with RR background fields which are still
not well understood.\footnote{For a preliminary 
discussion of some subleading ${\cal O}(1)$
contributions to anomalies see \cite{daak}.}

However, as pointed out in \cite{apty}, certain subleading $1/N$
corrections (i.e.\ terms of order $N$) in theories with open or
unoriented strings, corresponding to $SO(N)$ or $Sp(N)$ gauge theories,
may be accessible.  In \cite{apty} an ${\cal N}=2$ superconformal field
theory arising from D3-branes on a $\ZZ_{2}$ orientifold O7-plane with
D7-branes \cite{fts} was analyzed. In particular, in a rather subtle
analysis it was shown that the order $N$ contribution to the chiral $U(1)$
R-current anomaly, proportional to $(a-c)$,  is correctly reproduced in
the dual supergravity theory on $\adsx$, where $X^{5} = S^{5}/\ZZ_{2}$ 
\cite{sgd}, by bulk Chern-Simons couplings on the D7 and O7 world-volumes.

By supersymmetry, this chiral anomaly is related to the trace anomaly and
therefore indirectly \cite{apty} also confirms the AdS/CFT correspondence
for the conformal anomaly to this order in the large $N$ expansion. The
purpose of this note is to perform a direct calculation of the trace 
anomaly along the lines of \cite{hs}. By chasing the Chern-Simons
couplings from type I' theory back to ten dimensions, we see that
they originate from the Green-Schwarz couplings $H^{2}=(dB + \omega_{L} 
+\ldots)^{2}$ in the heterotic string. These terms are known to be
related by supersymmetry to CP even $R^{2}$-terms
proportional to the Riemann tensor squared and 
$F^{2}$-terms for the gauge group $SO(8)\ss SO(32)$ \cite{ghmr2,eric}.

By using heterotic - type I duality and 
T-dualizing to type I', we show that these terms
give rise to order $N$ Riemann tensor and gauge field strength
squared terms in eight dimensions
leading to a subleading order $N$ contribution to the conformal
anomaly upon reduction to $AdS_{5}$. 
We then show that the external gauge field contribution
and the crucial coefficient of the 
$\mbox{Riem}^{2}$-term of the boundary metric in the conformal anomaly, 
proportional to $(a-c)$, are precisely reproduced by the supergravity
calculation. 

We also find other terms of order $N$, proportional to the squares
of the Ricci tensor and Ricci scalar. This particular linear combination
differs from that of the field theory result precisely by a term 
of order $N$ attributable to an effective five-dimensional cosmological 
constant. We have been unable to determine this contribution
because of our ignorance regarding other four-derivative terms in the
type I' theory like $(F^{(5)})^{4}$ and $R(F^{(5)})^{2}$. Conversely,
comparing the supergravity calculation with the known field theory
result gives a concrete (but not in itself particularly interesting)
prediction for the $1/N$ contribution to the effective cosmological 
constant in this theory.

Another class of theories with subleading order $N$ corrections to the
trace anomaly are ${\cal N}=4$ SYM theories with orthogonal or symplectic
gauge groups. These can be realized as low-energy theories on D3-branes
at an orientifold O3-plane and a candidate for their supergravity dual is
tpye IIB string theory on an $AdS_{5}\times \RR\PP^{5}$ orientifold. At
first, these theories appear to present a puzzle as there are no
D-branes or O-planes wrapping the $AdS_{5}$ and therefore there can be
no orientifold or open string corrections to the bulk theory. We 
will show that both the leading and the subleading order $N$ 
contributions to 
the anomaly are correctly reproduced by the classical Einstein 
action by taking into account the (fractional) RR charge of the O3-plane.

In section 2, we review the CFT side of the gravitational and
external gauge field contributions to the conformal anomaly. 
In section 3, we deduce the relevant $R^{2}$ and $F^{2}$ 
terms in the $AdS_{5}$ supergravity action via heterotic - type I
- type I' duality. In section 4, we review the calculation of 
the leading ${\cal O}(N^{2})$ contribution to the trace anomaly 
following \cite{hs}. We then deduce the ${\cal O}(N)$
contributions by extending the analysis of \cite{hs} to include
the $R^{2}$ and $F^{2}$ terms (section 5). In section 6 we 
discuss the ${\cal N}=4$ SYM theories for SO and Sp gauge groups,
and we conclude with a discussion of the `missing'
contributions due to an effective cosmological constant of order $N$.

\section{The Trace Anomaly on the CFT Side}

The field theory of interest is \cite{apty} an ${\cal N}=2$
superconformal field theory with $Sp(N)$ gauge group, and 4 fundamental
and one antisymmetric traceless hypermultiplet. It arises \cite{fts} as the
low-energy theory on the world volume on $N$ D3-branes sitting 
inside eight D7-branes at an O7-plane. Among the global symmetries
of the theory there are an $SO(8)$-symmetry (from the D7-branes)
as well as an $SU(2)\times U(1)$ R-symmetry of the ${\cal N}=2$
superconformal algebra. Taking the near-horizon limit of this 
configuration one finds \cite{sgd} that the conjectural string 
theory dual of this theory is type IIB string theory on $\adsx$
where $X^{5}=S^{5}/\ZZ_{2}$ in which the D7 and O7 fill the 
$AdS_{5}$ and are wrapped around an $S^{3}$ which is precisely 
the fixed point locus of the $\ZZ_{2}$. Because of the $\ZZ_{2}$
action, the relation between the five-form flux $N$ and the 
curvature radius of $AdS_{5}$ is now
\be
L = (8\pi g_{s}N)^{1/4}\ell_{s}
\label{L2}
\ee
instead of (\ref{L1}). We will set $\ell_{s}=1$ in the following.

The trace anomaly, when the theory is coupled to an external metric, is
\be
\langle T^{\mu}_{\;\mu}\rangle = -a E_{4} - cI_{4}\;\;,
\ee
where, using shorthand notation, 
\be
\mbox{Riem}^{2}=R_{ijkl}R^{ijkl}\;\;,
\ee
with $R_{ijkl}$ the Riemann curvature tensor of the metric of the 
(boundary) space-time etc.,
\bea
E_{4}&=&\frac{1}{16\pi^{2}}\left( \mbox{Riem}^{2} - 4 \mbox{Ric}^{2}
+\mbox{R}^{2}\right)\non
I_{4}&=&-\frac{1}{16\pi^{2}}\left( \mbox{Riem}^{2} - 2 \mbox{Ric}^{2}
+\frac{1}{3}\mbox{R}^{2}\right)
\eea
Thus
\be
\langle T^{\mu}_{\;\mu}\rangle = \frac{1}{16\pi^{2}}
[(c-a)\mbox{Riem}^{2} +(4a-2c)\mbox{Ric}^{2} +(\trac{1}{3}c-a)\mbox{R}^{2}]
\;\;.
\ee
and we see that the $\mbox{Riem}^{2}$-term is proportional to $(c-a)$.

The coefficients $a$ and $c$ are determined in terms of the field content
of the theory. In particular, for the vector- and hypermultiplets
of the ${\cal N}=2$ theories one has
\bea
a_{V}=\frac{5}{24}&& c_{V}=\frac{1}{6}\non
a_{H}=\frac{1}{24}&& c_{H}=\frac{1}{12}
\eea
Thus, for one ${\cal N}=4$ multiplet ($n_{V}=n_{H}=1$) one has 
$a=c=1/4$ and the trace anomaly of ${\cal N}=4$ $SU(N)$ SYM theory
is
\be
\langle T^{\mu}_{\;\mu}\rangle = 
\frac{N^{2}-1}{32\pi^{2}}[\mbox{Ric}^{2}-\trac{1}{3}\mbox{R}^{2}]\;\;.
\label{tsun}
\ee
For an ${\cal N}=2$ theory with $n_{V}$ vector multiplets and $n_{H}$
hypermultiplets one has
\be
\langle T^{\mu}_{\;\mu}\rangle = \frac{1}{24\times 16\pi^{2}}
[(n_{H}-n_{V})\mbox{Riem}^{2} +12 n_{V}\mbox{Ric}^{2} -
\trac{1}{3}(11n_{V}+n_{H})\mbox{R}^{2}]\;\;.
\ee
In the present case, with
\bea
n_{V}&=&N(2N+1)=2N^{2}+N\non
n_{H}&=& 4\times 2N + N(2N-1) - 1 =  2N^{2} + 7N -1
\eea
one finds
\bea
a_{total}&\equiv& n_{V}a_{V} + n_{H}a_{H} = \frac{1}{24}(12N^{2}+12N -1)\non
c_{total}&\equiv& n_{V}c_{V} + n_{H}c_{H} = \frac{1}{24}(12N^{2}+18N -2)
\eea
and 
\be
a_{total}-c_{total} = \frac{1}{24}(1-6N)\;\;,
\ee
and the conformal anomaly is
\be
\langle T^{\mu}_{\;\mu}\rangle = \frac{1}{24\times 16\pi^{2}}
\left[
(6N-1) \mbox{Riem}^{2}+ (24N^{2}+12N)\mbox{Ric}^{2}
-(8N^{2}+6N-\frac{1}{3}) \mbox{R}^{2}\right]\;\;.
\ee

The leading ${\cal O}(N^{2})$ contribution is
\be
\frac{N^{2}}{16\pi^{2}}[\mbox{Ric}^{2}-\trac{1}{3}\mbox{R}^{2}]\;\;.
\label{leading}
\ee
This is exactly twice the ${\cal N}=4$ result and this is in accordance
with the expected \cite{hs,sg} relation between volumes, $\mbox{Vol}(S^{5})$ 
versus $\mbox{Vol}(X^{5})$, and the leading contribution to the anomaly. On
the supergravity side this term arises \cite{hs} from  a regularization
of the (divergent) classical gravity action which is just a volume
integral in this case as the $AdS$ scalar curvature (the Einstein-Hilbert
Lagrangian) is constant for $AdS_{5}$. We will review this calculation 
below.

The subleading ${\cal O}(N)$-term is
\be
\frac{6N}{24\times 16 \pi^{2}}[\mbox{Riem}^{2}+2 \mbox{Ric}^{2}-\mbox{R}^{2}]
\;\;.
\label{subleading}
\ee
We will show that, modulo undetermined volume terms of the form
(\ref{leading}) (with coefficients of order $N$ rather than $N^{2}$),
this term arises from a Riemann curvature squared term in the bulk gravity
action (with the precise numerical coefficient deduced from that appearing
in the heterotic string through heterotic - type I - type I' duality).

One can also couple the theory to external gauge fields of a flavour
symmetry group $G$. 
In general, the contribution of gauge fields to the trace anomaly
has been shown in \cite{cdj} to be proportional to the beta-function
of the corresponding gauge coupling constant. The result obtained
in \cite{cdj} is
\be
\langle T^{\mu}_{\;\mu}\rangle_{G} 
= \frac{\beta(g)}{2g} F^{a}_{ij}F^{a\,ij}\;\;,
\label{tbeta}
\ee
where $\beta(g)$ has the standard form
\be
\frac{\beta(g)}{2g}= -\frac{g^{2}}{32\pi^{2}}\left[\frac{11}{3}c_{2}(G)
-\frac{4}{3}T(R_{f}) - \frac{1}{3}T(R_{s}) \right]+{\cal O}(g^{4})\;\;.
\label{beta}
\ee
Here $R_{f,s}$ are the representations of $G$ on the (Dirac)
fermions and (complex) scalars respectively, and $T(R)$ is the
Dynkin index of the representation $R$, 
\be
\Tr_{R}t_{a}t_{b} = T(R)\d_{ab}\;\;.
\ee
To apply this result in the present situation we note the following.
First of all, in Euclidean space there is a minus sign on the right
hand side of (\ref{tbeta}), as can be seen by tracing through the
derivation in \cite[section 3]{cdj}. Moreover,
for an external gauge field, the first term on the right hand side of 
(\ref{beta}) is of course absent.
In the present case, we can choose $G=SO(8)$, and the only fields 
that are charged under $G$ are the $8N$ fundamental hypermultiplets
in the fundamental representation of $SO(8)$. As an ${\cal N}=2$
hypermultiplet in four dimensions consists of one Dirac fermion and
two complex scalars, the contribution of external $SO(8)$ gauge fields
to the trace anomaly is
\be
\langle T^{\mu}_{\;\mu}\rangle_{G} 
= -\frac{N T(\r{8})}{16\pi^{2}} F^{a}_{ij}F^{a\,ij}\;\;.
\label{TG}
\ee
We have dropped the factor $g^{2}$ because we will be working with
the scaled gauge fields in terms of which the action takes the form
$S=(1/4g^{2})\int F^{2}+\ldots$. 

We see that this term is also of order $N$, and we will show that
this contribution to the anomaly is reproduced precisely by an
$F^{2}$-term in the heterotic - type I action or, alternatively,
by the $F^{2}$-term of the Dirac-Born-Infeld D7-brane action
(wrapped on the $S^{3}\subset X^{5}$).  

\section{The $R^{2}$ and $F^{2}$ Terms}

In \cite{apty}, the relevant Chern-Simons terms in the $AdS_{5}$ bulk
action arose from terms proportional to 
\be
\int C^{(4)}\wedge \Tr (\Omega\wedge\Omega)\;\;,
\int C^{(4)}\wedge \Tr (F\wedge F)\;\;,
\ee
in the world volume theory of the D7-branes and O7-planes, where
$\Omega$ is the Riemann curvature two-form, $F$ denotes the 
$SO(8)$ gauge field and $C^{(4)}$ is the RR
4-form coupling to the D3-brane. T-dualizing these terms
to type I, the RR 4-form becomes a six-form $C^{(6)}$ coupling to 
the type I 5-brane. Writing this interaction as $F^{(7)}\wedge\omega_{L,YM}$,
where $\omega_{L,YM}$ is the Lorentz / Yang-Mills
Chern-Simons term, and dualizing
$F^{(7)}=*dB$, we see that this term arises from the 
modification 
\be
H^{2}= (dB + \la_{L} \omega_{L} -\la_{YM}\omega_{YM})^{2}
\ee
of the $B$-field kinetic term in the type I and heterotic supergravity 
actions required by the Green-Schwarz anomaly cancellation mechanism.

Now it is known
\cite{ghmr2,eric} that supersymmetry relates this term to a four-derivative
CP even term $R_{LMNP}R^{LMNP}$ in the ten-dimensional heterotic action 
together with the $SO(32)$ Yang-Mills term. The relevant part of the
heterotic action for our purposes is thus
\be
S_{h}=\frac{1}{16\pi(8\pi^{6})}
\int d^{10}x\sqrt{G^{h}}\ex{-2\phi^{h}}
(R + \frac{1}{4}(R_{LMNP}R^{LMNP} - F^{a}_{MN}F^{a\,MN}))\;\;.
\ee
We will now first check explicitly that these two terms in the end
give rise to terms of order $N^{2}$ and $N$ in the AdS supergravity
action respectively. We will determine the precise numerical factors 
below.

First of all, using the rules of heterotic - type I duality,
\bea
\phi^{h}&=&-\phi^{I}\non
G^{h}_{MN} &=& \ex{-\phi^{I}}G^{I}_{MN}\;\;,
\eea
in the type I theory one obtains
\be
\int d^{10}x\sqrt{G^{I}}(\ex{-2\phi^{I}} R + 
\frac{1}{4} \ex{-\phi^{I}} 
(R_{LMNP}R^{LMNP} - F^{a}_{MN}F^{a\,MN}))\;\;.
\ee
Notice that the Riemann tensor square term comes with $\exp (-\phi)$
rather than with $\exp(-2\phi)$. This indicates that it arises from
disc (D9-branes) and crosscap (orientifold O9-planes) world-sheets
and not from
the sphere. The latter was to be expected since the sphere calculation
is identical to that in IIB where one knows that there is no $R^{2}$-term.
The observation that type I - heterotic duality dictates the 
appearance of an $R^{2}$-contribution from the disc diagram in
type I theory was originally made in \cite{at}.

For a constant dilaton, which is all that we are interested in, the
dependence of the action on the type I string coupling constant 
$g_{I}$ (defined in general by $g_s=\exp<\phi_s>$,
for $s=h, I, I'$ respectively) is thus
\be
S_{I} \sim\int d^{10}x\sqrt{G^{I}}(\frac{1}{g_{I}^{2}}R 
+ \frac{1}{4} \frac{1}{g_{I}}(R_{LMNP}R^{LMNP} - F^{a}_{MN}F^{a\,MN}))\;\;.
\ee 
Now we T-dualize this on a two-torus of volume $V_{I}$ to type I' theory in 
eight dimensions. Since the eight-dimensional Newton constant is invariant, 
we have (modulo factors of $2$ and $2\pi$)
\be
V_{I}/g_{I}^{2} = V_{I'}/g_{I'}^{2}\sim 1/V_{I}g_{I}^{2}\;\;,
\ee
and therefore
\be
g_{I} \sim g_{I'}/V_{I'}
\ee
and 
\be
V_{I}/g_{I} \sim 1/g_{I'}\;\;.
\ee
Thus the T-dualized eight-dimensional action is
\be
S_{I'} \sim\frac{V_{I'}}{g_{I'}^{2}} 
\int d^{8}x\sqrt{G^{I'}} R + 
\frac{1}{4} \frac{1}{g_{I'}}
\int d^{8}x\sqrt{G^{I'}} (R_{LMNP}R^{LMNP}-F^{a}_{MN}F^{a\,MN})\;\;.
\ee
Since T-duality takes D9-branes to D7-branes and O9-planes to O7-planes,
one sees that in type I' the $R^{2}$- and $F^{2}$-terms come from discs 
attached to
the D7-branes and crosscaps corresopnding to O7-planes. This explains why
there is no transverse volume factor $V_{I'}$ in these terms. 

Now, to extract the $N$-dependence of these terms we scale the 
metric to unit radius, not forgetting to scale $V_{I'}$ as well. 
Thus 
\bea
V_{I'}&\ra& L^{2}V_{I'} \non
d^{8}x\sqrt{G^{I'}} &\ra& L^{8} d^{8}x\sqrt{G^{I'}} \non
R &\ra& L^{-2}R \non
R_{LMNP}R^{LMNP}&\ra& L^{-4}R_{LMNP}R^{LMNP}\;\;,
\eea
and the action becomes
\be
S_{I'} \sim\frac{L^{8}V_{I'}}{g_{I'}^{2}} 
\int d^{8}x\sqrt{G^{I'}} R + 
\frac{1}{4} \frac{L^{4}}{g_{I'}}
\int d^{8}x\sqrt{G^{I'}} (R_{LMNP}R^{LMNP}-F^{a}_{MN}F^{a\,MN})\;\;.
\ee
Using (\ref{L2}) we see that, as anticipated, the string coupling
constant drops out and the Einstein term is of order $N^{2}$ while
the curvature squared terms in the effective 7-brane action
are of order $N$. For a recent discussion of curvature squared
terms in type II D-brane actions see \cite{bmg}.

The precise numerical factors of the five-dimensional action can now also
be determined. For the Einstein term, plus the cosmological constant,
we have the inverse ten-dimensional Newton constant times the volume
$\mbox{Vol}(X^{5})=\mbox{Vol}(S^{5})/2$ times, as we have seen, $L^{8}$,
giving
\bea
S_{E}&=&\frac{1}{16\pi(8\pi^{6} g_{I'}^{2})}\times \frac{\pi^{3}}{2} 
\times (8\pi
g_{I'}N)^{2} \times \int_{AdS_{5}}d^{5}x \sqrt{G} (R-2\Lambda)\non
&=&\frac{N^{2}}{4\pi^{2}} \int_{AdS_{5}}d^{5}x \sqrt{G} (R-2\Lambda)\;\;,
\label{adsr}
\eea
where $R$ now denotes the five-dimensional Ricci scalar.

For the Riemann tensor squared term in five dimensions, and related terms
arising from the dimensional reduction of the internal and mixed 
components of this term, the numerical coefficient arises as follows.
There is a factor of 1/4 in the ten-dimensional action. It was related
by supersymmetry to the anomaly cancelling Green-Schwarz term for 
the gauge group $SO(32)$. By turning on appropriate Wilson lines in the
type I theory, this gauge group can be reduced to $SO(8)^{4}$. Upon
T-duality, these Wilson lines translate into the positions of the 
D7-branes in the type I' theory. 

Thus there are clusters of 8 D7-branes, each of the clusters located at
one of the 4 O7 orientifold planes. As we have seen above, the total
$R^{2}$-term comes from discs attached to D7-branes and crosscaps for
the O7-planes, each of the four clusters giving $\trac{1}{4}$ of the 
total contribution. As in the near horizon limit three of these clusters
are infinitely far away, only one quarter of this term 
will be relevant. 

Moreover, because of the presence of the orientifold,
the volume of the two-torus should be taken to be $(2\pi^{2})$ rather than
the usual $(2\pi)^{2}$. Wrapping the D7 branes on the $S^{3}$, the
fixed locus of the $\ZZ_{2}$ action, produces another contribution
$\mbox{Vol}(S^{3})$. This $S^{3}$ has \cite{sgd,apty} the standard volume 
$2\pi^{2}$. Finally, there is, as we have seen above, a factor of
$L^{4}$ from the scaling of the metric cancelling the $1/g_{I'}$. 

Putting everything together,
we find that the coefficient of the $R^{2}$-term (as well as
that of the other components of this term and other related 
4-derivative terms in the action) is 
\bea
S_{R^{2}} &=&\frac{1}{16 \pi \times 8\pi^{6}}\times 
(8\pi N) \times 2\pi^{2} \times 2\pi^{2} \times \frac{1}{16}
\times 
\int_{AdS_{5}}d^{5}x \sqrt{G}R_{\mu\nu\rho\sigma}R^{\mu\nu\rho\sigma} +
\ldots\non
&=& \frac{6N}{24 \times 16 \pi^{2}}
\int_{AdS_{5}}d^{5}x \sqrt{G}R_{\mu\nu\rho\sigma}R^{\mu\nu\rho\sigma} +
\ldots
\label{adsr2}
\eea
Note the striking similarity of this coefficient with the subleading 
contribution (\ref{subleading}) to the trace anomaly. Even though we
haven't even begun to calculate the contribution of this term to the
trace anomaly, this certainly suggests that we are on the right track. 

The same argument shows that the $\Tr F^{2}$-term for the
$SO(8)\subset SO(8)^{4}\subset SO(32)$ gauge fields in the heterotic
action, reinstating the factor of 4 we divided by before,
gives rise to an order $N$ contribution
\be
S_{F^{2}} 
= -\frac{N}{16 \pi^{2}}
\int_{AdS_{5}}d^{5}x \sqrt{G} F^{a}_{\mu\nu}F^{a\,\mu\nu} + \ldots
\label{adsf2}
\ee
to the bulk action. 
Alternatively \cite{apty}, up to an overall
normalization, the coefficient of this
term could have been deduced from the $SO(8)$ Dirac-Born-Infeld action 
of the D7-branes. From this point of view it is of
course obvious that this is an open string disc contribution and hence of
order $N$. 
The relative factor of 4 between the gravitational and gauge field
couplings mirrors that found in \cite{apty} for the five-dimensional
Chern-Simons terms arising from the D7/O7 RR Chern-Simons couplings.

Note again the striking similarity of this term with
the contribution (\ref{TG}) of external $SO(8)$ gauge fields to the
trace anomaly. Once we have developed the appropriate machinery below,
it will be straightforward to verify that (\ref{adsf2}) reproduces
exactly the anomaly (\ref{TG}). 

\section{Review of the ${\cal O}(N^{2})$ Calculation}

\subs{The Strategy}

Before embarking on the calculation of the ${\cal O}(N)$ contribution
to the trace anomaly, let us quickly review the calculation of the
leading ${\cal O}(N^{2})$ contribution \cite{hs}. 

Because the AdS metric has a second order pole at infinity, AdS space
only induces a conformal equivalence class $[\gz_{ij}]$ of metrics
on the boundary.  To check for conformal invariance, one chooses a
representative $\gz_{ij}$ acting as a source term for the energy-momentum
tensor of the boundary theory. The AdS/CFT correspondence predicts that
the CFT effective action in the large $N$ supergravity limit is
\be
W_{CFT}(g_{0})=S_{grav}(g;g_{0})\;\;,
\ee
where $S_{grav}(g;g_{0})$ denotes the gravitational action evaluated on a
classical configuration which approaches (in the conformal sense) the metric
$\gz_{ij}$ on the boundary. The action is the sum of two terms, the standard
bulk action $S_{E} \sim\int (R-2\Lambda)$, and a boundary term, involving
the trace of the extrinsic curvature of the boundary, required
to ensure the absence of boundary terms in the variational principle. 
To solve the classical equation of motion
\be
R_{\mu\nu} - \trac{1}{2}g_{\mu\nu}(R-2\Lambda)=0\;\;,
\label{eins}
\ee
with this boundary condition, one can \cite{graham,hs} make the following
ansatz for the metric,
\be
G_{\mu\nu}dx^{\mu}dx^{\nu} = \frac{L^{2}}{4}\frac{d\rho^{2}}{\rho^{2}}
+ \frac{1}{\rho} g_{ij}dx^{i}dx^{j}\;\;,
\label{adsg}
\ee
with the boundary sitting at $\rho=0$.  We will set $L=1$ in the following
as our scaling arguments use the unit radius metric.
The metric $g_{ij}$ has an expansion as \cite{graham,hs}
\be
g_{ij} = g^{(0)}_{ij} + \rho g^{(2)}_{ij} + \rho^{2}g^{(4)}_{ij} 
+\rho^{2}\log\rho h^{(4)}_{ij} + \ldots\;\;,
\label{grho}
\ee
with $\gz_{ij}$, as above, the chosen boundary value. 

Now, for a solution to the classical equations of motion, both the bulk 
and the boundary term are divergent (the former because for an Einstein
manifold the classical Einstein-Hilbert action reduces to a volume 
integral, and the latter because the induced metric on the boundary is
singular). Therefore, one needs to regularize this expression (which, in
view of its conjectured relation to the CFT effective action is not
surprising). This can be done by introducing a cutoff $\epsilon$  
restricting the range of $\rho$ to $\rho \geq \epsilon$. Note that,
in agreement with general arguments on holography \cite{lsewapjp}, 
this bulk IR cutoff corresponds to an UV cutoff in the CFT. Then the
regularized CFT effective action $W^{\epsilon}_{CFT}(\gz)$ is invariant
under $\d\gz = \la\gz, \d\epsilon = \la\epsilon$. 

$W^{\epsilon}_{CFT}(\gz)$ can be written
as a sum of terms diverging as $\epsilon\ra 0$, $W^{\infty}_{CFT}(\gz)$,
and a finite term $W^{fin}_{CFT}(\gz)$. The former is a sum of terms
which are integrals of local covariant expressions in the boundary
metric $\gz$ and hence they can be removed by local counterterms. 
Among these terms there is, for $AdS_{d+1}$ with $d$ even, a logarithmically
divergent term (which, interestingly enough, does not arise from the
logarithmic term in the expansion (\ref{grho}) of the metric).
In the standard way, removal of this term will then
induce a conformal anomaly in the finite part $W^{fin}(\gz)$. 
The boundary term never contributes to the conformal anomaly (this is
a consequence of the fact that the logarithmic
term $h^{(4)}_{ij}$ in (\ref{grho}) is known to be traceless 
with respect to $\gz$ \cite{graham}) and we will not consider it in the 
following. 

\subs{Calculation of the ${\cal O}(N^{2})$ Contribution}

In the case at hand, the precise form of the anomaly is determined
as follows. For an Einstein space with $R_{\mu\nu}=-4 g_{\mu\nu}$, the
value of the classical Lagrangian is $L_{c}=-8$. 
The volume element is
\be
\sqrt{\det G}= \trac{1}{2}\rho^{-3}\sqrt{\det g}\;\;,
\label{detG}
\ee
where the latter can be expanded as
\be
\sqrt{\det g} = \sqrt{\det g^{(0)}}(1+\trac{1}{2}\rho \Tr g^{(2)}
+\trac{1}{8} \rho^{2}[(\Tr g^{(2)})^{2}-\Tr ((g^{(2)})^{2})])+\ldots\;\;.
\label{detg}
\ee
Here, $\Tr$ denotes the trace with respect to the metric $g^{(0)}$ and 
we have made use of the useful identity \cite{hs}
\be
\Tr g^{(4)} = \trac{1}{4}\Tr ((g^{(2)})^{2})\;\;.
\label{hs}
\ee
By iteratively solving the Einstein equations as a power series
in $\rho$, one finds \cite{hs}
\be
\gt_{ij} =- \trac{1}{2}(r^{(0)}_{ij}-\trac{1}{6}\gz_{ij}r^{(0)})\;\;.
\label{hs2}
\ee
Here $r^{(0)}_{ij}$ denotes the Ricci tensor of $\gz$ etc. Note that
we are using the opposite sign conventions of \cite{hs}. Our conventions
for the curvature tensor,
\be
R^{\la}_{\;\sigma\mu\nu}=
\del_{\mu}\G{\la}{\sigma}{\nu}
-\del_{\nu}\G{\la}{\sigma}{\mu}
+\G{\la}{\mu}{\rho}\G{\rho}{\nu}{\sigma}
-\G{\la}{\nu}{\rho}\G{\rho}{\mu}{\sigma}\;\;,
\ee
and the Ricci tensor, 
\be
R_{\mu\nu} := R^{\la}_{\;\mu\la\nu} = g^{\la\sigma} R_{\sigma\mu\la\nu}
\;\;,
\ee
are such that the curvature of the sphere is positive. 

We will need the square of the trace and the trace of the square of this
term. One has
\bea
\Tr \gt &=& -\trac{1}{6}r^{(0)}\non
(\Tr \gt)^{2} &=& \trac{1}{36}(r^{(0)})^{2}\non
\Tr (\gt)^{2} &=& \trac{1}{4}( r^{(0)}_{ij} r^{(0)ij}-
\trac{2}{9}(r^{(0)})^{2})\;\;.
\eea
In particular, therefore, the order $\rho^{2}$-term in the expansion
(\ref{detg}) is 
\be
(\Tr\gt)^{2}-\Tr(\gt)^{2} = -\frac{1}{4}
[r^{(0)}_{ij}r^{(0)ij}- \frac{1}{3}(r^{(0)})^{2}]\;\;.
\label{volume}
\ee

As one obtains a $\rho^{-3}$ from $\sqrt{G}$, it is clear that a
logarithmically divergent term will arise only from the term of order
$\rho^{2}$ in (\ref{detg}). In particular, we see that for any
gravitational action including only the Einstein term and a cosmological
constant, the leading contribution to the conformal anomaly will be 
proportional to (\ref{volume}). Comparing with the discussion in 
section 2, we see that this impies $a=c$ to order $N^{2}$ as (\ref{volume})
does not contain a $\mbox{Riem}^{2}$-term. 

Let us now apply this to $\adsx$ and thus to the leading 
contribution to the trace anomaly of the ${\cal N}=2$ superconformal
field theory considered in \cite{apty} and above. 
Using (\ref{adsr}), and noting that the factor of $1/2$ in (\ref{detG})
is cancelled by a conventional factor of 2 in the definition of the 
conformal anomaly, one finds that the ${\cal O}(N^{2})$ conformal anomaly, 
i.e.\ the coefficient of the $\log\epsilon$-term, is
\bea
&&\frac{N^{2}}{4\pi^{2}} \times (-8) \times \frac{1}{8} 
\times [(\Tr g^{(2)})^{2}-\Tr ((g^{(2)})^{2})]\non
&=& \frac{N^{2}}{16\pi^{2}} 
[r^{(0)}_{ij}r^{(0)ij}- \frac{1}{3}(r^{(0)})^{2}]\;\;.
\eea
This is indeed {\em precisely} the leading contribution (\ref{leading}) 
to the conformal anomaly calculated on the CFT side.

\section{The ${\cal O}(N)$ Contribution}

\subs{The Strategy}

Now the strategy for including the Riemann tensor sqared
term should be clear. We take the Einstein plus Riemann squared action
(\ref{adsr}) plus (\ref{adsr2}) (possibly also with the $F^{2}$-term
(\ref{adsf2}) - we will comment on the inclusion of this term below)
\be
S=\frac{N^{2}}{4\pi^{2}} \int d^{5}x \sqrt{G} (R-2\Lambda)
+\frac{6N}{24\times 16 \pi^{2}}
\int d^{5}x \sqrt{G}R_{\mu\nu\rho\sigma}R^{\mu\nu\rho\sigma} + \ldots
\label{news}
\ee
(plus boundary terms),
solve the equations of motion with the
given boundary metric $\gz$, and isolate the log-divergent terms in the
action evaluated on this classical solution. Note that, because of
the presence of the term $R^{\mu\nu\rho\sigma}R_{\mu\nu\rho\sigma}$ 
in (\ref{news}) this calculation will no longer reduce to just a volume 
calculation. 

In principle, of course, solving the classical equations of motion 
of this higher-derivative gravity action to the required order in
$\rho$ is an unpleasant task. In the present case, however, a 
drastic simplification is brought about by the fact that we are only
interested in the contributions of order $N$ to the classical action.
For this, it is sufficient to evaluate the term
$R^{\mu\nu\rho\sigma}R_{\mu\nu\rho\sigma}$ on the classical solution
of the previous section to the original Einstein equation (\ref{eins}).

Indeed, as the second term in (\ref{news}) is $1/N$ down with respect to
the Einstein term, we can make an ansatz for the solution to the 
full equations in the form 
\be
G_{\mu\nu}=G_{\mu\nu}^{(0)} + \frac{1}{N}G^{(1)}_{\mu\nu}\;\;,
\ee
where $G^{(0)}_{\mu\nu}$ is a solution of (\ref{eins}). Plugging this
solution into the Einstein term, i.e.\ the first term of (\ref{news}),
one obtains at order $N^{2}$ the leading contribution to the anomaly
calculated in the previous section. A term of order $N$ that could 
potentially arise as the next term in the expansion 
is actually zero (because we are expanding about a
classical solution to the Einstein action and the boundary term is
precisely there to cancel any residual boundary terms). The second
term in (\ref{news}) will give a contribution of order $N$ when 
evaluated on $G_{\mu\nu}^{(0)}$, and any other contributions
involving $G_{\mu\nu}^{(1)}$ will be of order 1 or lower. 

Therefore, to find the order $N$ contributions to the trace anomaly,
we need to 
\begin{enumerate}
\item calculate the Riemann curvature tensor of the metric 
(\ref{adsg}), with $g_{ij}$ given by (\ref{grho}) and (\ref{hs2}),
as a function of $\rho$, and 
\item then determine the order $\rho^{-1}$-terms in the 
$\rho$-expansion of 
\be
\sqrt{\det G}G^{\a\mu}G^{\b\nu}G^{\ga\la}G^{\d\sigma}
R_{\a\b\ga\d}R_{\mu\nu\la\sigma}\;\;.
\ee
\end{enumerate}
Since we know that $\sqrt{\det G} \sim \rho^{-3}\sqrt{\det g}$ (\ref{detG}),
this means that we need to pick up the order $\rho^{2}$-terms from
\be
\sqrt{\det g}G^{\a\mu}G^{\b\nu}G^{\ga\la}G^{\d\sigma}
R_{\a\b\ga\d}R_{\mu\nu\la\sigma}\;\;,
\ee
Here the $\rho$-expansions of the curvature, of $\sqrt{\det g}$ and of
the inverse metric have to be considered.\footnote{Similar calculations have
recently also been performed in \cite{NO}, however with the diametrically 
opposite motivation of trying to reproduce the leading ${\cal O}(N^{2})$
contribution to the anomaly from a higher derivative action \ldots\ldots.} 

If one also includes the $F^{2}$-term (\ref{adsf2}), then in principle
one would of corse have to solve the coupled Einstein-Yang-Mills equations.
But as the $F^{2}$-term is also of order $N$ the same argument as above 
shows that the resulting subleading corrections to the metric are again
irrelevant. As regards the equation of motion for $F$ itself, we will see
below that only the boundary value of $F$ contributes so we do not have to
solve these equations either. 

\subs{External Gauge Fields}

Let us begin with the external gauge field contribution (\ref{TG}) to
the anomaly as it is by far the simplest contribution to determine
(much simpler, in fact, than even the leading ${\cal O}(N^{2})$ contribution
to the anomaly discussed above). 

As the above discussion shows, we need
to pick up the order $\rho^{2}$-terms of
\be
\sqrt{\det g} G^{\a\mu}G^{\b\nu}F^{a}_{\a\b}F^{a}_{\mu\nu}\;\;.
\ee
Now the components of the inverse metric are $G^{\rho\rho}=4\rho^{2}$,
$G^{ij}=\rho g^{ij}$, where $g^{ij}$ has the expansion
\be
g^{ij}= g^{(0)ij} -\rho g^{(2)ij} 
+ \rho^{2}(((g^{(2)})^{2})^{ij}-g^{(4)ij}) + \ldots,
\ee
where indices are raised with $g^{(0)ij}$. As the two inverse metrics
contribute at least a factor of $\rho$ each, the only contribution
to the anomaly arises from 
\be
\sqrt{\det g^{(0)}} g^{(0)ik}g^{(0)jl}  F^{(0)a}_{ij}F^{(0)a}_{kl}\;\;,
\ee
where $F^{(0)}_{ij}$ is the boundary value of the gauge field 
\be
F_{ij} = F^{(0)}_{ij} + {\cal O}(\rho)\;\;.
\ee
In \cite{apty} the relation betwwen the bulk supergravity and boundary
SCFT $SO(8)$-generators, in the fundamental representation $\r{8}$, was
determined from the AdS/CFT correspondence. Using this result,
one obtains that  $T(\r{8})$, appearing in the field theoretic
expression (\ref{TG}), is equal to 1.  Therefore,
(\ref{adsf2}) gives precisely the 
external gauge field contribution (\ref{TG})
to the trace anomaly.

Alternatively, this term could have been deduced (in the Abelian,
non-interacting case) by following the prescription in \cite{ew1}:
On-shell, the bulk Maxwell action reduces to a boundary term, and 
this boundary term can be evaluated in terms of Witten's bulk-to-boundary
Green's functions, extracting the local term (relevant to the anomaly)
in the end. 

More directly, one can proceed locally, i.e.\ without using Green's
functions, by solving the 
Maxwell equations in a $\rho$-expansion as was done for the Einstein
equations in \cite{hs}. From this vantage point, the logarithmic
divergence arises directly in the boundary term $\sim \int
A_{i}\del_{\rho}A_{i}$ because a term of order $\rho\log\rho$ in the
$\rho$-expansion of $A_{i}$ turns out to be
required to solve the bulk Maxwell 
equations (cf.\ the $\rho^{2}\log\rho$-term in the expansion (\ref{grho})
of the metric, required for the same reason).

\subs{The Curvature Tensor}

As three different metrics appear here, $G_{\mu\nu}$, $g_{ij}$ and
$g^{(0)}_{ij}$, we will correspondingly denote their 
curvature tensors by $R^{\mu}_{\;\nu\la\rho}$, $r^{i}_{\;jkl}$, 
$r^{(0)i}{}_{jkl}$.  $\nabla$ will denote the covariant derivative 
compatible with $g_{ij}$, $\nabla^{(0)}$ that compatible with $g^{(0)}_{ij}$.
$\rho$-derivatives will be denoted by a prime.

The ubiquitous combination $g_{ij}-\rho g'_{ij}$, which we will abbreviate to
$k_{ij}$ in the following, contains no terms linear in 
$\rho$. Up to $\rho^{2}\log\rho$-terms one has
\be
k_{ij} \equiv g_{ij}-\rho g'_{ij} = g^{(0)}_{ij}-\rho^{2}(g^{(4)}+h^{(4)}) 
+ \ldots
\ee
We will sometimes also abbreviate $g^{(4)}+h^{(4)}=f^{(4)}$. 


For the curvature tensor one then finds
\bea
R_{ijkl} &=& \rho^{-1}[r_{ijkl}+
\rho^{-1} (k_{il}k_{jk}-k_{ik}k_{jl})]\non
R_{\rho ijk} &=& \trac{1}{2}\rho^{-2}(\nabla_{j}k_{ik}-\nabla_{k}k_{ij})\non
R^{\rho}_{\;i\rho j}&=& -2\rho^{-1}(k_{ij}-\rho k'_{ij})
+\rho^{-1} k^{2}_{ij}\;\;.
\eea
where in the last line the product is taken with respect to the metric
$g_{ij}$. 

We will need the $\rho$-expansion of these curvature tensors.
For $R_{ijkl}$, we need to expand $r_{ijkl}$ as well as the other
terms. Symbolically we have
\be
r^{i}_{\;jkl} = r^{(0)i}_{\;jkl} + \rho
(\nabla^{(0)}_{k}\d\G{i}{j}{l}-\nabla^{(0)}_{l} \d\G{i}{j}{k}) + \ldots,
\ee
where we do not need to know the precise form of the $\d\Ga$'s.
Using this and the definition of $k_{ij}$ one finds
\bea
R_{ijkl} &=& G_{in}R^{n}_{\;jkl} = \rho^{-1}g_{in}R^{n}_{\;jkl}\non
R_{ijkl} &=& \rho^{-2}[\gz_{il}\gz_{jk}-\gz_{ik}\gz_{jl}]\non
         &+& \rho^{-1} r^{(0)}_{ijkl}\non
         &+& \rho^{0} [ \gz_{ik}\ff_{jl} +\gz_{jl}\ff_{ik}
                       -\gz_{il}\ff_{jk} -\gz_{jk}\ff_{il}]\non
         &+& \rho^{0} [\nabla^{0}_{k}\d\Ga_{ijl}-\nabla^{(0)}_{l} \d\Ga_{ijk}]
\non
         &+& \rho^{0} [\gt_{in}r^{(0)n}_{\;jkl}] + {\cal O}(\rho)
\label{rijkl}
\eea

$R_{\rho ijk}$ is simpler, we just keep the first term (and not even that
one will contribute as we will see),
\bea
R_{\rho ijk} &=&
\rho^{-1}[-\trac{1}{2}(\nabla_{j}\gt_{ik}-\nabla_{k}\gt_{ij})]
+ {\cal O}(1)
\eea

For $R^{\rho}_{\;i\rho j}$, one has
\bea
R^{\rho}_{\;i\rho j}&=& \rho ^{-1}[-\gz_{ij}]\non
                    &+& \rho^{0} [-\gt_{ij}]\non
                    &+& \rho^{+1}[-5 \ff_{ij} + (\gt)^{2}_{ij}]+ {\cal
O}(\rho^{2})\;\;,
\label{rij}
\eea
where now, of course, in the last line the product is taken with respect
to $\gz$.

As mentioned above, we need to pick up the order $\rho^{2}$-terms from
\be
\sqrt{\det g}G^{\a\mu}G^{\b\nu}G^{\ga\la}G^{\d\sigma}
R_{\a\b\ga\d}R_{\mu\nu\la\sigma}\;\;,
\ee

Let us deal with $R_{\rho ijk}$ first. In that case, the factor 
entering the contractions is
\be
\sqrt{\det g}G^{\rho\rho}G^{im}G^{jn}G^{kp}\;\;.
\ee
This will contribute at least $\rho^{2} \times \rho^{3} = \rho^{5}$,
but the highest negative power of $\rho$ that can arise from the 
square of $R_{\rho ijk}$ is $\rho^{-2}$, giving an overall $\rho^{3}$
and therefore no contribution to the anomaly. 

For $R_{ijkl}$ we have
\be
\sqrt{\det g}G^{im}G^{jn}G^{kp}G^{lq}\;\;.
\label{volijkl}
\ee
This contributes $\rho^{4}$ and higher powers. But the highest negative
power arising from $R_{ijkl}^{2}$ is $\rho^{-4}$. Hence here terms of
order $\rho^{4}$, $\rho^{5}$ and $\rho^{6}$ in the expansion of
the contraction/volume factor (\ref{volijkl}) are relevant. 
At order $\rho^{n}$, $n=4,5,6$ respectively, one has:
\bea
\rho^{4}:&& \sqrt{\gz} g^{(0)im} g^{(0)jn} g^{(0)kp} g^{(0)lq} \non
\rho^{5}:&& \sqrt{\gz} \trac{1}{2}\Tr\gt 
g^{(0)im} g^{(0)jn} g^{(0)kp} g^{(0)lq} \non
&-&4 \sqrt{\gz} g^{(2)im} g^{(0)jn} g^{(0)kp} g^{(0)lq} \non
\rho^{6}:&& \sqrt{\gz}\trac{1}{8}[(\Tr\gt)^{2}-\Tr(\gt)^{2}] 
g^{(0)im} g^{(0)jn} g^{(0)kp} g^{(0)lq} \non
&-&4 \sqrt{\gz} \trac{1}{2}\Tr\gt g^{(2)im} g^{(0)jn} g^{(0)kp} g^{(0)lq} \non
&+& 4\sqrt{\gz} [((\gt)^{2}-\gf)^{im}] g^{(0)jn} g^{(0)kp} g^{(0)lq} \non
&+& 2\sqrt{\gz} g^{(2)im} g^{(2)jn} g^{(0)kp} g^{(0)lq} \non
&+& 4\sqrt{\gz} g^{(2)im} g^{(0)jn} g^{(2)kp} g^{(0)lq} 
\label{ijkl}
\eea

Finally, for $R^{\rho}_{\;i\rho j}$, the structure is
\be
\sqrt{\det g}G^{im}G^{jn}\;\;.
\label{volij}
\ee
This will be of order $\rho^{2}$ and higher. On the other hand, the 
square of the curvature tensor gives terms of order $\rho^{-2}$ and 
higher. Hence in the  expansion of the contraction/volume factor
(\ref{volij}), 
terms of order $\rho^{2}$, $\rho^{3}$, $\rho^{4}$ are relevant. These
are
\bea
\rho^{2}:&& \sqrt{\gz} g^{(0)im} g^{(0)jn} \non
\rho^{3}:&& \sqrt{\gz} \trac{1}{2}\Tr\gt 
g^{(0)im} g^{(0)jn} \non
&-&2 \sqrt{\gz} g^{(2)im} g^{(0)jn} \non
\rho^{4}:&& \sqrt{\gz}\trac{1}{8}[(\Tr\gt)^{2}-\Tr(\gt)^{2}] 
g^{(0)im} g^{(0)jn} \non
&-&2 \sqrt{\gz} \trac{1}{2}\Tr\gt g^{(2)im} g^{(0)jn} \non
&+& 2\sqrt{\gz} [((\gt)^{2}-\gf)^{im}] g^{(0)jn} \non
&+& \sqrt{\gz} g^{(2)im} g^{(2)jn} 
\label{ij}
\eea

\subs{Contributions from $R_{ijkl}$}

Let us call the five contributions in (\ref{rijkl}) $I$, $II$, $III$, 
$IV$ and $V$. 
Three terms contribute to the $\rho^{4}$-term of (\ref{ijkl}), namely
$II\times II$, $I\times III$ and $I\times V$. 
$I\times IV$ only contributes a total
derivative of a covariant quantity and can therefore be cancelled 
by the variation of a local counterterm. Using the tracelessness
of $h^{(4)}$ one sees that $h^{(4)}$
will not contribute either. The other terms give
\bea
\rho^{4}, II\times II&& r^{(0)}_{ijkl} r^{(0)ijkl}\non
\rho^{4}, I \times III&& -6 \Tr (\gt)^{2}\non
\rho^{4}, I \times V&& -4 \gt_{ij} r^{(0)ij}
\eea
where we have used (\ref{hs}). The two terms of order $\rho^{5}$
in (\ref{ijkl}) need to be paired with $I\times II$: 
\bea
\rho^{5}, I\times II&& -2 r^{(0)}\Tr\gt +16 g^{(2)ij} r^{(0)}_{ij}
\eea
The terms of order $\rho^{6}$ in (\ref{ijkl}) need to be paired with
$I\times I$. From the first three terms of order $\rho^{6}$ we get
\bea
\rho^{6}, I\times I&& -9 (\Tr\gt)^{2} + 15 \Tr (\gt)^{2}
\eea
The fourth and fifth term give 
\bea
\rho^{6}, I\times I&& 4 (\Tr\gt)^{2} -4 \Tr (\gt)^{2}\non
                   && 4 (\Tr\gt)^{2} +8 \Tr (\gt)^{2}
\eea
Adding all this up, we find the subtotal from $R_{ijkl}$ to be
\bea
&& r^{(0)}_{ijkl} r^{(0)ijkl}
-2 r^{(0)}\Tr\gt +12 g^{(2)ij} r^{(0)}_{ij}
- (\Tr\gt)^{2} + 13 \Tr (\gt)^{2}\non
&=&
 r^{(0)}_{ijkl} r^{(0)ijkl}-\frac{11}{4}r^{(0)}_{ij}r^{(0)ij}
+\frac{7}{12}(r^{(0)})^{2}\;\;.
\eea

\subs{Contributions from $R^{\rho}_{\;i\rho j}$}

We proceed as above. The three terms of (\ref{ij}) we call
$I$, $II$, $III$. Every contribution has to be multiplied by 
four, because there are four components of the Riemann tensor
with two $\rho$'s.
\bea
\rho^{2}, II\times II&& 4\Tr(\gt)^{2}\non
\rho^{2}, I \times III&& 2 \Tr (\gt)^{2}\non
\rho^{3}, I \times II&& 
4(\Tr\gt)^{2}-16 \Tr (\gt)^{2}
\eea
There are four terms of order $\rho^{4}$ in (\ref{ij}), to be paired
with $I\times I$. These give
\bea
\rho^{4}, I \times I&&  2(\Tr\gt)^{2}-2 \Tr (\gt)^{2}\non
                    &&  4 \Tr (\gt)^{2}\non
                    &&  6 \Tr (\gt)^{2}\non
                    &&  -4(\Tr\gt)^{2}
\eea
Adding all these up, one gets
\be
2(\Tr\gt)^{2}-2\Tr (\gt)^{2}=- \frac{1}{2}r^{(0)}_{ij} r^{(0)ij}
+ \frac{1}{6}(r^{(0)})^{2}
\;\;.
\ee

\subs{The Total ${\cal O}(N)$ Contribution to the Trace Anomaly}

Adding up all the above contributions, and remebering the prefactor
in (\ref{adsr2}), we find that supergravity
predicts the ${\cal O}(N)$ contribution to the trace anomaly to 
be
\be
 \frac{6N}{24 \times 16 \pi^{2}}\times[
 r^{(0)}_{ijkl} r^{(0)ijkl}-\frac{13}{4}r^{(0)}_{ij}r^{(0)ij}
+\frac{3}{4}(r^{(0)})^{2}]\;\;.
\ee
A glance at (\ref{subleading}) shows that this does not yet look
particularly encouraging. However, let us split these terms as
\bea
&& \frac{6N}{24 \times 16 \pi^{2}}\times 
[ r^{(0)}_{ijkl} r^{(0)ijkl} +2 r^{(0)}_{ij}r^{(0)ij}
 - (r^{(0)})^{2}]\non  
&-& \frac{6N}{24 \times 16 \pi^{2}} \times \frac{21}{4}
[r^{(0)}_{ij}r^{(0)ij}- \frac{1}{3}(r^{(0)})^{2}]\;\;.
\eea
We see that the first term reproduces {\em precisely} the subleading
contribution (\ref{subleading}) to the conformal anomaly, in particular
with the crucial term proportional to $(a-c)$. The
second (error) term, on the other hand, is exactly (and this is an 
important check on our calculation) of the form of a volume contribution 
(\ref{volume}), just like the leading ${\cal O}(N^{2})$-term. We will
say more about the possible origin of this volume term below.

\section{${\cal O}(N)$ Corrections for SO and Sp ${\cal N}=4$ Theories}

In this section we will briefly discuss another class of models which,
at first sight, seems to present a puzzle. Looking back at the ${\cal
N}=4$, $SU(N)$ trace anomaly (\ref{tsun}), we see that there is a
tree-level contribution, determined in \cite{hs}, no term of order
$N$ but an ${\cal O}(1)$ correction that ought to arise from a string
one-loop calculation. For other gauge groups $G$, however, the situation
is different.  In general, one has
\be
\langle T^{\mu}_{\;\mu}\rangle =
\frac{\mbox{dim}(G)}{32\pi^{2}}[\mbox{Ric}^{2}-\trac{1}{3}\mbox{R}^{2}]\;\;.
\label{tson}
\ee
In particular, for orthogonal (symplectic) gauge groups $SO(N)$ ($Sp(N/2)$
for $N$ even), 
$\mbox{dim}(G)$ contains both quadratic and linear terms in $N$,
\bea
\mbox{dim}(SO(2k)) &=& k(2k-1)\non
\mbox{dim}(SO(2k+1)) = \mbox{dim}(Sp(k)) &=& k(2k+1)\;\;,
\label{dimG}
\eea
and we want to understand the origin of these linear terms in the AdS/CFT
correspondence. 

${\cal N}=4$ theories with gauge groups $SO(N)$, $Sp(N/2)$ 
can be realized as the low-energy dynamics of $N$ parallel
D3-branes at an orientifold O3-plane \cite{jp}, i.e.\
with the branes sitting at the singularity of a transverse 
$\RR^{6}/\ZZ_{2}$. Here $\ZZ_{2}$ acts as $\vec{x}\ra -\vec{x}$ for
$\vec{x}\in\RR^{6}$. Note that, because of the non-compactness of
the transverse space, the number of D3-branes is not fixed by 
RR tadpole cancellation.

This strongly suggests \cite{ew2} that a string
theory dual to these theories is given by type IIB string theory
on an $AdS_{5}\times \RR\PP^{5}$ orientifold. 

Clearly, unlike for the ${\cal N}=2$ theory we discussed above, now there are 
no branes wrapping the entire $AdS_{5}$. 
So where are the terms linear in $N$ going to come
from? The answer is: from the classical Einstein action itself. The
reason for this is that O$p$-planes themeselves are carriers of 
RR-charge, and hence the numerical value of $N$ appearing in the classical
D$p$-brane solutions (in the coefficient of the term $r^{p-7}$ in the
corresponding harmonic function, $r$ being the transverse distance
from the brane),
will be shifted in the presence of an orientifold
O$p$-plane. In particular, O3-planes carry fractional RR charge
$\pm\frac{1}{4}$ \cite{jp,mukhi}. With the minus sign, one obtains
$SO(N)$, and with the plus sign, for $N$ even, $Sp(N/2)$
gauge theories. In fact, at least prior to taking 
the near-horizon limit, in the coefficient of the leading order
${\cal{O}}((\frac{\sqrt{\alpha'}}{r})^{7-p})$ correction to the classical
D$p$-brane solution, this $N$-independent term arises from the addition 
of the crosscap $\RR\PP^{2}$ orientifold contribution (of order $g_s$)
to the disc D-brane diagram (of order $g_s N$). 

This should extend to all orders to reproduce the
expected result so that, in the AdS-limit, 
the net-effect of the presence of
the O3-plane is to replace $N$ as appearing in (\ref{L2}) by 
\be
N \ra \frac{N}{2} \pm\frac{1}{4}\;\;,
\ee 
where we took also into account that only $N/2$ of the $N$ 
D3-branes lie on $\RR^{6}/\ZZ_{2}$. Consequently
\be
L^{4} = 8\pi g_{s}\left(\frac{N}{2}\pm\frac{1}{4}\right)\;\;.
\ee
Repeating the calculation in (\ref{adsr}) and 
section 4 for the leading contribution
to the trace anomaly, we now find
\bea
\langle T^{\mu}_{\;\mu} \rangle &=& 
\frac{1}{16\pi \times 8\pi^{6}g_{s}^{2}}
\times \mbox{Vol}(\RR\PP^{5}) \times L^{8} \times 
\int_{AdS_{5}}d^{5}x \sqrt{G} (R-2\Lambda)\non
&=& 
\frac{1}{16\pi \times 8\pi^{6}}
\times \frac{\pi^{3}}{2} \times 64 \pi^{2} \times (\frac{N}{2}
\pm\frac{1}{4})^{2} \times
\frac{1}{4}[\mbox{Ric}^{2}-\trac{1}{3}\mbox{R}^{2}]\non
&=& \frac{1}{32\pi^{2}}\frac{N(N\pm 1)+\frac{1}{8}}{2}
[\mbox{Ric}^{2}-\trac{1}{3}\mbox{R}^{2}] \;\;.
\eea
Comparing with (\ref{dimG}), we see that to order
$N$ this agrees exactly with the trace anomaly formula 
(\ref{tson}) for $G=SO(N)$ and $G=Sp(N/2)$. In these cases
we have therefore been able to reproduce both the leading and
the subleading order
$N$ corrections directly from the classical Einstein 
action by taking 
into account the fractional RR charge of the O3-plane.

\section{Discussion}

We have shown that supergravity calculations with higher-derivative
actions are capable of reproducing the subleading corrections to 
the CFT trace anomaly. In this particular example, on the basis
of the results of \cite{apty} this was to be expected on general grounds
since supersymmetry relates the chiral and trace anomalies. Nevertheless,
we find it quite remarkable that the somewhat messy (even though
straightforward) classical calculations performed above conspire to give
precisely the correct result for the trace anomaly in the end.

As regards the `missing' volume contribution, we have of course
attempted to determine this in a variety of ways but it seems 
to us that a definitive answer requires a better understanding 
of $\a'$-corrections and supersymmetrization of the type I'
effective action. 

As the expansion (\ref{detg}) produces $1/8$ times
(\ref{volume}), we see that what we are missing is an effective
cosmological constant term 
\be
- \frac{168 \times 6N}{24 \times 16 \pi^{2}}\int_{AdS_{5}}\sqrt{G}d^{5}x
\;\;.
\ee
There are many possible terms that can contribute to this cosmological
constant. For instance, we have so far neglected the contributions of
the internal and mixed components of $R_{LMNP}R^{LMNP}$. These 
contributions can in principle be determined either from duality arguments
or by a direct two-point function calculation on the type I' side. 
There may also be four-derivative terms of the metric involving 
the squares of the Ricci tensor or Ricci scalar (perhaps in the form
of the familiar Gauss-Bonnet combination). Such terms are afflicted
by the usual field redefinition ambiguities. Finally, there may also
be terms involving $(F^{(5)})^{4}$ and mixed terms of the 
type $R(F^{(5)})^{2}$. The former correspond to four-point functions
on the type I' side and  a direct calculation of these terms, although
possible in principle, is somewhat cumbersome.

One might have hoped to be able to invoke heterotic - type I - type I'
duality once more to fix these terms. For instance, in the heterotic
string it is known \cite{eric} that supersymmetry forces the 
CP-even four-derivative terms involving $g_{MN}$ and $B_{MN}$ to appear
as curvature-squared terms of the connection with torsion $H=dB + \ldots$.
As $H$ eventually dualizes to $F^{(5)}$ in the type I' theory, this is
the sort of restrictive structure one might have hoped for. However,
chasing these terms through the dualities is somewhat problematic. 

For one, as one is taking a large volume limit on the type I' side,
this corresponds to a small two-torus on the type I side, and thus
it appears that winding mode contributions in Type I need also be
considered. Also, at a purely classical level, in order to T-dualize
the D7/O7/D3 configuration underlying the ${\cal N}=2$ theory 
we have been considering to a type I confiiguration of D9/D5 branes 
one needs to delocalize it in the transverse directions. But if one
does that, it will no longer have the same near-horizon limit
(T-duality and near-horizon limits do not commute). 
Conversely, we have been unable to find a classical type I solution
which gives the desired configuration on the type I' side and which
could have been used to calculate the effective cosmological constant
directly on the type I side.

All this just confirms the general picture that appears to be emerging
from the work done on the AdS/CFT correspondence, namely that 
whatever can be checked reliably confirms the conjectured
correspondence, but that even simple (one-loop, anomaly) field theory 
calculations are difficult to reproduce on the AdS side. Clearly,
what is required among other things is a better understanding of
string theory with RR backgrounds.

\subsubsection*{Acknowledgements}

We are grateful to George Thompson and Hossein Sarmadi for 
useful discussions at various stages of this work. This 
work was supported in part by the EC under the TMR contract 
ERBFMRX-CT96-0090. 

\rnc{\Large}{\normalsize}

\end{document}